\documentclass[conference]{IEEEtran}
\IEEEoverridecommandlockouts
\usepackage{cite}
\usepackage{amsmath,amssymb,amsfonts}
\usepackage{algorithmic}
\usepackage{graphicx}
\usepackage{textcomp}
\usepackage{xcolor}
\usepackage{amsmath}
\usepackage{amsthm}
\usepackage{etoolbox}
\usepackage{hyperref}

\def\BibTeX{{\rm B\kern-.05em{\sc i\kern-.025em b}\kern-.08em
    T\kern-.1667em\lower.7ex\hbox{E}\kern-.125emX}}
\newcommand{\insertfig}{
\begin{center}
    \includegraphics[width=0.85\linewidth, keepaspectratio]{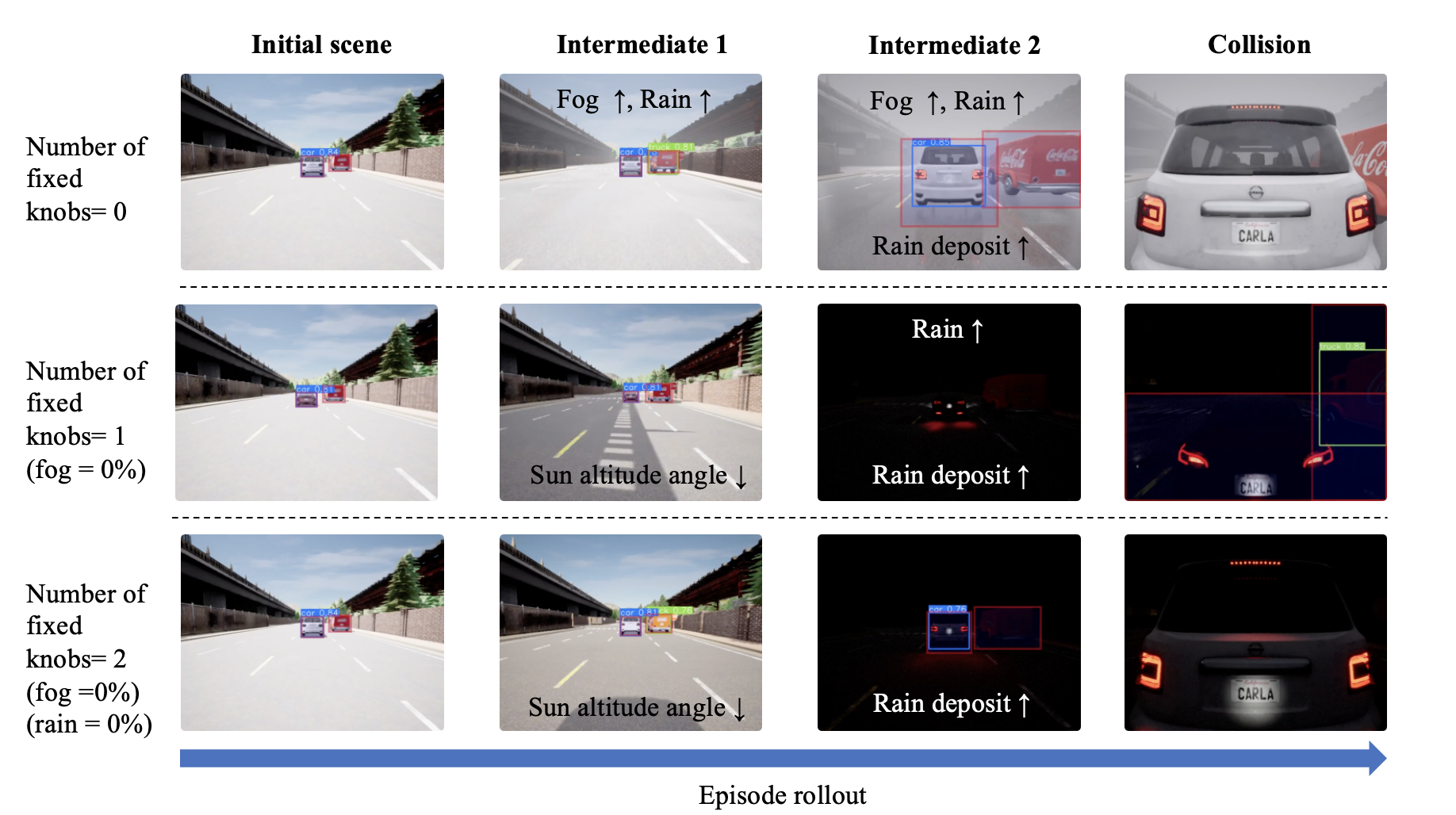}\\
    {\small GENESIS-RL: A reinforcement learning framework to progressively manipulate the environment (weather conditions in this \\example) for an autonomous system (autonomous vehicle in this case) to systematically synthesize natural edge cases that \\may lead to system-level safety issues (collision in this case). Project website: \href{https://hjy77.github.io/GENESIS-RL/}{https://hjy77.github.io/GENESIS-RL/}}
\end{center}}
\makeatletter
\apptocmd{\@maketitle}{\centering\insertfig}{}{}
\begin{document}

\newcommand{\rulebook}{\mathcal{\Lambda}}
\newcommand{\rules}{\Lambda}
\newcommand{\realizations}{\Sigma}
\newcommand{\reals}{\mathbb{R}}
\newcommand{\nnreals}{\reals_{\geq 0}}
\newcommand{\nok}[1]{\textcolor{red}{#1}}

\newtheorem*{remark}{Remark}
\title{GENESIS-RL: GEnerating Natural Edge-cases with Systematic Integration of Safety considerations and Reinforcement Learning\\
\thanks{This work was partly supported by the National Science
Foundation, USA under grants CNS-1845969, CNS-2141153, CNS-1954556.}
}

\author{
    \IEEEauthorblockN{
        Hsin-Jung Yang\IEEEauthorrefmark{1}, 
        Joe Beck\IEEEauthorrefmark{2}, 
        Md Zahid Hasan\IEEEauthorrefmark{1},
        Ekin Beyazit\IEEEauthorrefmark{1}\\
        Subhadeep Chakraborty\IEEEauthorrefmark{2}, 
        Tichakorn Wongpiromsarn\IEEEauthorrefmark{1},
        Soumik Sarkar\IEEEauthorrefmark{1}
    }
    \IEEEauthorblockA{
        \IEEEauthorrefmark{1}Iowa State University, Ames, IA, USA\\ \IEEEauthorrefmark{2}University of Tennessee, Knoxville, TN, USA\\
        Email: \{hjy, zahid, ekin, nok, soumiks\}@iastate.edu, \{jbeck9, schakrab\}@utk.edu
    }
}
\maketitle

\begin{abstract}
In the rapidly evolving field of autonomous vehicles, the safety and reliability of the system components are fundamental requirements. These components are often vulnerable to complex and unforeseen environments, making natural edge-case generation essential for enhancing system resilience. This paper presents GENESIS-RL, a novel framework that leverages system-level safety considerations and reinforcement learning techniques to systematically generate naturalistic edge cases. By simulating challenging conditions that mimic the real-world situations, our framework aims to rigorously test entire system's safety and reliability. Our experimental validation, conducted on high-fidelity simulator underscores the overall effectiveness of this framework.
\end{abstract}


\section{Introduction}

Scenario-based testing is one of the key approaches for the validation of autonomous systems, especially those that incorporate learning-enabled components that are known to be susceptible to rare, unexpected (potentially out-of-distribution) scenarios~\cite{machines5010006}. 
This testing approach is vital not only for ensuring the safety and reliability of these systems but also for enabling them to identify and rectify potential failures in diverse, unforeseen situations. In this context, identifying and preparing for challenging or edge-case scenarios becomes critical. Synthesizing realistic edge-case samples and incorporating these into the training process~\cite{karunakaran2023mdpi, Zheng_2020_CVPR} can significantly enhance the resilience of learning-enabled modules against adversarial conditions. 
By exposing the learning modules to these pessimistic samples, systems gain the opportunity to learn from challenging data and better generalize across a spectrum of real-world conditions. 
However, given the vast amount of possible scenarios, manual creation of every scenario is infeasible, making automated edge-case generation crucial for scalability and effectiveness~\cite{wang2021advsim}.

In this regard, traditional adversarial attacks on machine learning models explore the vulnerability of the models by injecting imperceptible noise into the inputs~\cite{Deng2020advattack, yoon2023learning}. These input perturbation methods, while effective in degrading the model performance, typically generates unnatural and unrealistic samples, diverging from genuine real-world scenarios. Furthermore, these approaches usually target specific components of an autonomous system rather than assessing the system as a whole. This narrow focus can overlook the holistic behavior of the system, where, for instance, a failure in the perception module might be compensated by the system's control mechanisms, thus not leading to a failure at the system level. On the other hand, if the control system is not able to compensate, a relatively small error in the perception module may lead to a catastrophic system-level failure. This highlights the limitation of focusing solely on component-level vulnerabilities without considering the integrated operation of the entire system.

Generative models have been used to synthesize edge cases that are more realistic~\cite{joshi2019semantic}. However, they are known to produce samples with artifacts that compromise their realism. These models, including generative adversarial networks (GANs)~\cite{acgan2017}, diffusion models~\cite{chen2023advdiffuser}, and more recently, text-to-image generation models such as DALLE~\cite{dalle2021}, CogView~\cite{ding2021cogview}, can suffer from issues such as unnatural distributions~\cite{yoon2021adversarial}, distinct artifacts and unstable training~\cite{brock2018large, dhariwal2021diffusion}, and slow inference rates, limiting their effectiveness in producing realistic and natural scenarios~\cite{song2021scorebased, zhang2023fast}. 

In this paper, we aim to alleviate these challenges, by performing edge-case generation with system-level safety objectives while maintaining the naturalness of the generated scenarios. We employ the rulebook formalism~\cite{Censi:2019:Liability} to precisely specify system-level safety objectives and leverage the capabilities of Reinforcement Learning (RL) to guide the generation of scenarios that not only challenge the system across all its components but also resemble real-world conditions closely. By focusing on the end-to-end vulnerability of autonomous systems, our approach aims to generate scenarios where the system fails to adhere to rulebook safety rules, thereby identifying potential systemic failures. Also, our proposed framework ensures that the generated scenarios are not only challenging but also devoid of unrealistic artifacts (via use of high-fidelity simulators), offering a more effective and comprehensive approach to testing and validating the safety and reliability of autonomous systems. 

In summary, the key contributions of this paper are as follows:
\begin{itemize}
    \item We propose a synthetic edge case data generation framework for system-level safety concerns in learning-enabled autonomous systems.

    \item We propose an RL-based intelligent sampling technique that can identify parametric settings of high-fidelity simulators to generate natural edge cases that may lead to violation of safety rules by a learning-enabled autonomous system. 

    \item We provide extensive experimental validation of our framework using the CARLA simulator~\cite{dosovitskiyCARLAOpenUrban2017b}. We also demonstrate that a pre-trained RL policy can generate edge-cases for new scenarios with minimal to no training steps, thus accelerating the process of assessment and verification of learning-enabled autonomous systems. 
    
\end{itemize}

\section{Related Works}
Recent research has explored diverse approaches to generating edge cases. Efforts using cost functions to pinpoint high-risk scenarios have shown potential yet often neglect critical factors like unpredictable trajectories~\cite{chen2021adversarial, ding2020learning, ding2021multimodal}. Perception-based techniques, such as constant norm-based perturbation, target the system's perception capabilities but may not address the system's overall performance comprehensively~\cite{madry2018norm, karunakaran2023mdpi, joshi2019semantic, mukherjee2022generative}. While innovative, methods that extract and recreate accidents from videos face challenges in accurately replicating real-world complexity~\cite{xinxin2020csg}.

Additionally, some edge-case generation software toolkits, like VerifAI~\cite{verifai-cav19}, are capable of analysis, falsification, and data augmentation for systems utilizing ML architectures. These toolkits leverage an "abstract feature space" of higher-level information compared to the low-level "concrete feature space" of image pixels to search for rule violation scenarios in a given environment. Domain randomization effectively bridges the sim-to-real gap~\cite{muratore2021data, niu2021dr2l}, but it can lead to training an overly conservative policy depending on the range of randomization. System identification~\cite{du2021auto} offers a feasible solution by estimating environmental parameters through limited interaction with real-world scenarios.

Lastly, Bayesian optimization-based methods often generate challenging scenarios with limited diversity and insufficient complexity~\cite{karunakaran2023mdpi, chen2021adversarial}. These methods typically produce short scenario segments with limited environment interactions since they require a predefined parameter range~\cite{Birkemeyer2023}. Consequently, they limit the assessment of system performance and fail to capture realistic edge cases with diverse interactions.

\section{Background}

In this work, a \textit{system} refers to the entity that is being evaluated for its ability to navigate and perform tasks within variable conditions. It could be an autonomous vehicle or any computational model. The \textit{world} denotes the simulated surroundings in which the system operates, a construct designed to emulate real-world dynamics where every aspect can have an effect on the system's behavior. Lastly, an \textit{actor} is an entity other than the system that also lives in the world.

As an example, in an autonomous driving context, the system could be the ego vehicle, and the world is where the ego vehicle is situated. 
Other entities, such as other vehicles and pedestrians on the street, are actors. 
Together with crucial factors such as weather and road conditions (including road markings and traffic signs) that are not part of the system but could affect the system's behavior, they are all parts of the world. 



\subsection{Deep Reinforcement Learning}
Deep Reinforcement Learning (DRL)~\cite{mnih2013playing} is an extension of RL that harnesses the representational power of deep neural networks. At its core, DRL adheres to the Markov Decision Process (MDP) framework, mathematically formulated as a 4-tuple $(S, A, P, R)$, where:
\begin{itemize}
    \item $S$ represents the state space, comprising all conceivable states $s_t$ at a given time $t$. 
    \item $A$ denotes the action space, encompassing all actions $a_t$ available to the agent at time $t$.
    \item $P$ the transition probability function, indicating the probability of transitioning from one state $s_{t}$ to another state $s_{t+1}$ given an action $a_t$.
    \item $R: S \times A \times S' \rightarrow \reals$ is the reward function, which assigns a numerical reward for each transition between states under specific actions.
\end{itemize}

In a typical DRL setup, \textit{the DRL agent} is the entity that we hope to train, whereas \textit{the environment} is the setting or domain wherein the DRL agent operates and makes decisions, which encompasses all aspects mentioned above in the MDP framework, including the state space $S$, action space $A$, the transition probability function $P$ and rewards $R$.

Under this MDP framework, the agent's decision-making strategy at any time $t$ is governed by a policy $\pi$, which maps the current state $s_t$ to an action $a_t$. In DRL, this policy is represented with neural networks, denoted as $\pi_{\theta}$, where $\theta$ represents the neural network's trainable parameters. This configuration enables the agent to dynamically refine its strategy by updating $\theta$, thus improving its performance and adaptability in navigating the environment.

The objective of DRL is to discover an optimal policy $\pi_{\theta}^*$ that guides the agent to maximize the expected return along a trajectory $\tau$, which is a sequence of states and actions $(s_0, a_0, s_1, a_1, ..., s_T, a_T)$. The expected return is calculated as $J(\theta) = \mathbb{E}_{\tau \sim \pi_\theta}[\sum_{t=0}^{T} \gamma^t R(s_t, a_t, s_{t+1})]$, where $\gamma$ is the discount factor and $T$ the length of the trajectory. The concept of episodes emerges naturally from this setup. An episode describes a complete trajectory from an initial state to a terminal state~\cite{rl_sutton_barto}.

\subsection{Rulebook}

We will use the rulebook formalism \cite{Censi:2019:Liability} to precisely describe the correct behavior of the system. A rulebook consists of a set $\rules$ of rules; each is evaluated over realizations. A realization is defined as a sequence of states of the system and all the other actors and features in the world. Given a set $\realizations$ of realizations, a rule is defined as a function $\lambda : \realizations \to \nnreals$ that measures the degree of violation of its argument. If $\lambda(x) < \lambda(y)$, then the realization $y$ violates the rule $\lambda$ to a greater extent than does $x$. In particular, $\lambda(x) = 0$ indicates that a realization $x$ is fully compliant with the rule. Note that the definition of the violation metric might be analytical, ``from first principles'', or be the result of a learning process.

In this work, we utilize the rulebook to calculate the rewards. A higher violation score leads to increased rewards, encouraging the agent to explore scenarios that challenge the system's safety protocols and resilience, thereby generating critical edge cases. For a detailed description of the rules used for reward calculation, please refer to the \textit{Reward calculator} subsection in the \textit{Experiments} section.

\section{Methodology}

At a high level, GENESIS-RL utilizes DRL to dynamically explore and manipulate the conditions of a simulated world, aimed at generating challenging yet naturalistic edge-cases for a system. To achieve this, we parameterized the world with \textit{parametric knobs}—adjustable settings that control various aspects of the simulation, which in the case of autonomous driving, could include dynamic weather patterns, object placements, traffic flow, and so on. By adjusting these knobs, the DRL agent is provided with the capability to systematically probe and alter the simulated world, effectively simulating different edge cases that the system under test might encounter. 
\begin{remark}

Our objective is to craft and manipulate the world (via simulation) to induce challenging scenes. By doing so, we seek to generate edge cases that test the limits of the system's current capabilities, aiming to reveal potential failure cases. In contrast to typical DRL works, we do not focus on improving the system's capabilities in this work.
\end{remark}




\subsection{DRL Problem Formulation}
Following the MDP framework, we define the state space, action space, and reward of our problem as follows:
\subsubsection{State space} The state space encompasses all conceivable states $s_t$, including permutations of parametric knobs, the system's behaviors, other actors, and features of the world. This state representation captures the dynamics of the world and the DRL agent's action inputs, and is conveyed through information obtained by the system. 





\subsubsection{Action space}  The action space is the set of all possible actions $a_t$ available to the agent, corresponding to the adjustments the agent can make to the parametric knobs within the simulation. To ensure that the changes introduced by the DRL agent lead to scenes that are natural and realistic, we imposed constraints on the extent of modifications possible at each step. Specifically, we limit the maximum percentage change that can be applied to any parametric knob by the DRL agent in a single action. This measure prevents extreme, unrealistic variations in conditions, thereby maintaining the realistic nature of the simulated scenes while still challenging the system under test.

\subsubsection{Reward}  The reward mechanism is designed to motivate the DRL agent to discover edge cases. It comprises two components: the learning module loss $r_m$ and the violation score $r_v$ derived from the rulebook. The learning module loss is the loss experienced by the learning-enabled module within the system, which acts as a direct reward to the agent, where a lower loss indicates better performance of the module at performing its designated task. The violation score is an indirect reward provided to the agent due to the imperfection of the learning-enabled modules. For example, in autonomous driving, the rulebook evaluates the ego vehicle's trajectory against a set of predefined rules, penalizing actions that lead to unsafe scenarios. The total reward $r_t$ at time step $t$ is calculated as a combination of these two elements. 

\subsection{GENESIS-RL Framework}
To implement our DRL formulation, we designed a framework consisting of the following components: the DRL agent, the initial scene generator, the simulator, the system, and the reward calculator. The latter four together form an environment for the DRL agent, facilitating continuous learning of the DRL agent through dynamic interaction.

\subsubsection{DRL agent}
The DRL agent is the decision-making core. At each time step $t$, it obtains the current state $s_t$ of the environment and executes an action $a_t$.  The environment then responds to this action by evolving to a new state based on the updated parametric knobs of the simulated world and issues a scalar reward $r_t$ to the agent as feedback.

\subsubsection{Initial scene generator}
The initial scene generator is responsible for creating a distribution of the initial scenes (a configuration of physical objects, the system and actors) and sampling from them in the simulated world. It dictates the initial conditions the system will encounter, therefore determining the initial scene observed by the DRL agent. 

\subsubsection{Simulator}
The simulator provides a realistic and interactive backdrop where the DRL agent's actions and the system's outputs are executed and new frames are updated, reflecting the changes in real-time.

\subsubsection{System}
As defined in the background section, The system is the entity being evaluated within variable conditions.

\subsubsection{Reward calculator}
As defined in the previous section, the reward calculator calculates the reward $r_t$ for time step $t$.

\subsection{Training the DRL Agent}
\begin{figure}[htb]
\centering
\includegraphics[width=\linewidth]{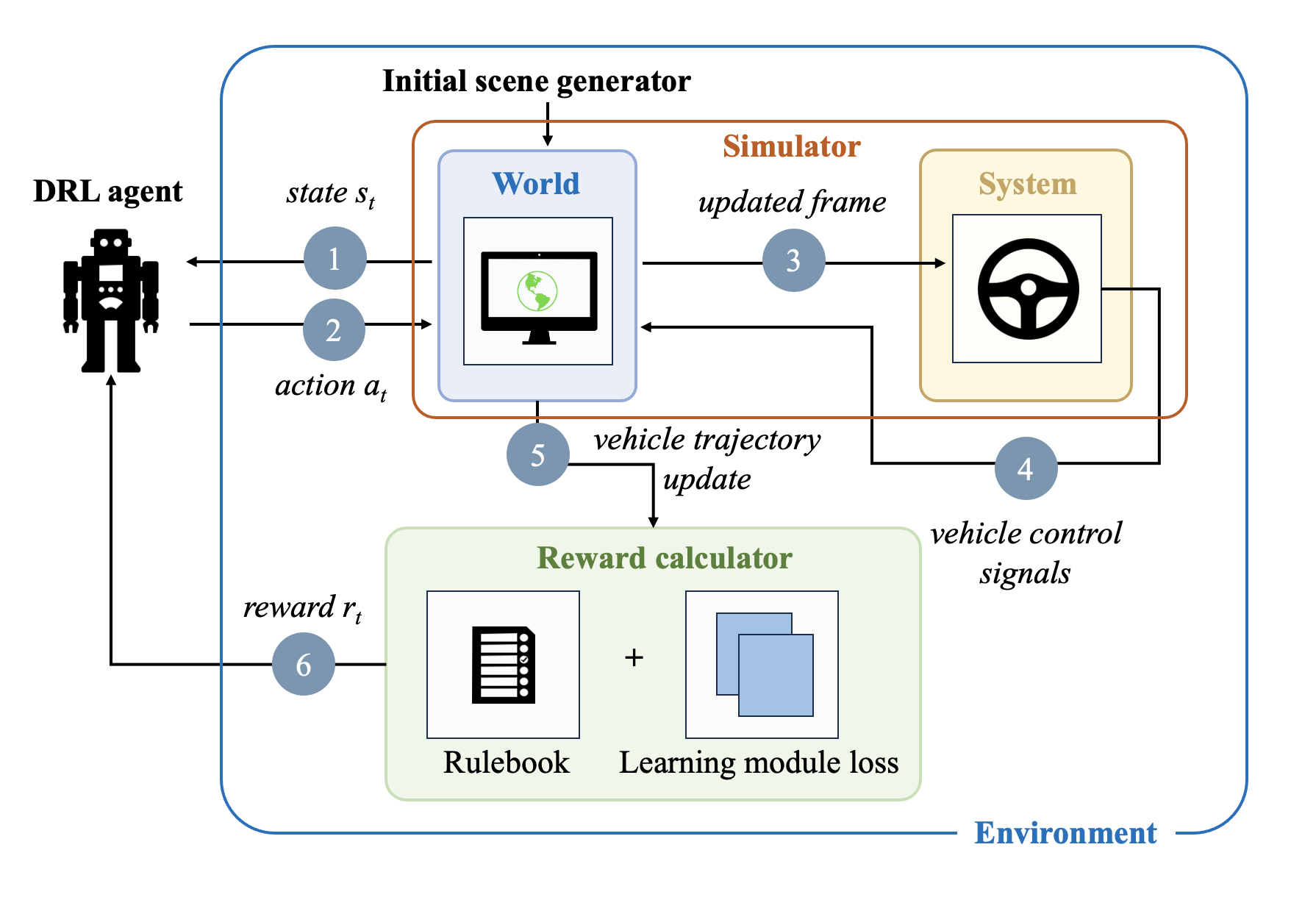}
\caption{Architectural overview of the proposed framework. At each step $t$, the DRL agent observes a state $s_t$ (1) and executes an action $a_t$ (2). The simulator then updates the simulated world accordingly and creates an updated frame. The updated frame (3) is then processed by the system to generate vehicle control signals (4). The control signals are then applied to the simulated world to update the vehicle trajectory (5). The reward calculator evaluates the performance by comparing the ego vehicle's trajectory against the rulebook and also computes the learning module loss, issuing a scalar reward $r_t$ (6) that guides the DRL agent's learning process.}
\label{framework}
\end{figure}

Putting things together, a single step of the training looks like as follows (See Fig.~\ref{framework}): at each time step $t$, the DRL agent receives a state $s_t$ from the simulator and executes an action $a_t$ on the simulator. The simulator reflects the changes based on the updated parametric knobs and the changes are subsequently captured by the system through its sensors. The system then generates control signals based on its inputs, which leads to a system trajectory update. The updated trajectory is then evaluated by the rulebook for violation score calculation and is sent back to the DRL agent combined with the learning module loss.

The DRL agent is trained through interactions with the environment, where it observes the states, applies actions, and receives rewards. The training process involves iterative episodes of simulation, during which the agent refines its policy $\pi_{\theta}$ to maximize the cumulative reward, effectively learning to identify and create challenging scenarios for the system.


\section{Experiments}
In this paper, we explore the weather conditions that can lead to natural edge cases for autonomous driving. Hence, we grant the DRL agent exclusive control over the weather conditions in the simulated world. The system we evaluate is the ego vehicle, tasked to navigate through the simulated world based on sensor feedback. In subsequent sections, we detail the operationalization of the GENESIS-RL framework's components, starting with the DRL agent and encompassing the environment, then the training and testing setups and evaluation metrics.

\begin{remark}
 Despite only having weather parameters as available actions in this work, we can also include other factors such as the behavior of other actors into our framework by parameterizing their behavior in the simulation (such as the aggressiveness of the other actors using CARLA's traffic manager, or the position and orientation of the actors using Scenic~\cite{scenic2019}). 
\end{remark}

\subsection{DRL Agent}
We have chosen the Proximal Policy Optimization (PPO) algorithm~\cite{schulman2017proximal} as the DRL agent for its stabilized training capabilities and proficiency in handling continuous state and action spaces. We adopted the implementation of Stable-Baselines3~\cite{stable-baselines3}. The state and action spaces are as follows:
\begin{itemize}
    \item State space: A 640$\times$480 three channel numpy array from the RGB camera attached to the ego vehicle. 
    \item Action space: A 5-tuple that represents the delta in value for each parametric knob. The parametric knobs are: the fog density, the precipitation density, the precipitation deposit level, the sun altitude angle, and the sun azimuth angle. These actions are bounded between $[-1, 1]$, and are scaled by a set of preset scalar factors to satisfy the perturbation limit, which is 5\%. For example, if the fog density ranges from 0 to 100, then the corresponding parametric knob would be scaled by a factor of 5 such that the change in fog density level between each step in an episode will not exceed 5\%.
\end{itemize}
As mentioned earlier, the DRL agent is developing a policy $\pi_{\theta}$ parameterized by a neural network $\theta$. Here, the architecture of $\theta$ is a CNN, which takes the RGB three channel arrays $(640, 480, 3)$ as inputs, and outputs a 5-tuple action.

\subsection{Environment}
The initial scene generator, the simulator, the system, and the reward calculator collaboratively create the environment of the DRL agent. These parts together generate the state and reward required to train the DRL agent and react to the actions of the DRL agent dynamically. 

\subsubsection{Simulator}
We employ CARLA~\cite{dosovitskiyCARLAOpenUrban2017b} to simulate intricate, dynamic urban settings with high visual fidelity.

\subsubsection{Initial scene generator}
Scenic~\cite{scenic2019} was utilized for creating the initial scenes as a strategic approach to ensure precision and versatility in our experimental setup. It allows us to craft realistic, detailed and specific initial scenes, thereby enhancing the relevance and challenge of each test instance presented to the system. Specifically, this generator is tasked with assigning the positions of vehicles within the simulator: For each initial scene, we introduce 30 vehicles, including the ego vehicle, a lead vehicle, a neighboring vehicle on the front right, and the remaining 27 vehicles randomly positioned within a specific radius of the ego vehicle. Vehicle colors, models, and the distance between the ego, lead, and neighbor vehicles could be either varied randomly or deterministic, depending on the use case. All vehicles except for the ego vehicle are set to CARLA's autopilot in order to provide a dynamic environment. 

We leveraged CARLA's~\cite{dosovitskiyCARLAOpenUrban2017b} record feature to efficiently catalog the configurations of the initial scenes into json files. These saved scenes are then randomly sampled and loaded back into the simulation between training or testing episodes. A total of 1199 initial scenes for \textit{Town05} and 1202 for \textit{Town10} were generated and cataloged using this generator.


\subsubsection{System}
The system in our study is an ego vehicle that navigates itself within the simulated world with RGB and depth cameras onboard. It is equipped with a perception-based controller, which consists of two parts: a perception model for object detection, utilizing a pre-trained YOLOv5s model~\cite{yolov5} to process RGB images from the onboard RGB camera, and a modified CARLA~\cite{dosovitskiyCARLAOpenUrban2017b} behavior controller that translates detection results into vehicle control signals. In our implementation, the input to the modified behavior controller includes the detection output from the perception model along with depth information from the onboard depth camera. By overlapping the 2D bounding boxes with the depth image, we were able to obtain the depth of the center point of the bounding boxes, which will then be used to obtain the 3D bounding boxes of the detected objects. Leveraging these 3D bounding boxes, combined with real-time data on the vehicle’s position, orientation, velocity, and the surrounding map, the controller adeptly synthesizes these inputs to generate vehicle control signals, i.e., throttle, brake, and steering signals.

\subsubsection{Reward calculator}
The reward $r_t$ is calculated by the weighted sum of the learning module loss $r_m$ and the violation score $r_v$. The learning module loss is defined by the Intersection Over Union (IOU) metric, assessing the system's object detection precision in real-time. Simultaneously, the violation score is generated by the rulebook evaluating a trajectory $\tau=s_1, s_2, s_3,....s_T$, consisting of the following two violation scores based on two critical safety rules: 
\begin{itemize}
    \item Vehicle collision rule, $\lambda_c(\tau)$, violated if the ego vehicle collides with other vehicles:
    \begin{align}
        \lambda_c(\tau) = \sum_{t=1}^T \alpha_c(s_t)
    \end{align}
    \begin{align}
    \centering
            \alpha_c(s_t) = v_c \text{ if collision happens, }
                0 \text{ otherwise}
    \end{align}
    \item Vehicle proximity rule, $\lambda_p(\tau)$, violated when the distance $d$ between the ego vehicle and the vehicle in front is less than a predetermined distance (which is set to 5 meters):
    \begin{align}
        \lambda_p(\tau) = \sum_{t=1}^T \alpha_p(s_t)
    \end{align}
    \begin{align}
    \centering
            \alpha_p(s_t) = v_p \text{ if $d$ $<$ 5}, 0 \text{ otherwise}
    \end{align}
    
\end{itemize} 
where $v_c$ and $v_p$ are the speeds of the ego vehicle in meters per second when a violation occurs. $\alpha_c(s_t)$ and $\alpha(s_t)$ are then weighted by their respective weights $w_c$ and $w_p$ then summed and transformed using the natural logarithm to compute $r_v$. Finally, the reward $r_t$ is calculated as follows:
\begin{align}
    r_t = r_m + r_v = e^{-iou} + ln(1+w_c\alpha_c+w_p\alpha_p)
\end{align}
where $iou$ is the IOU metric, $(w_c, w_p) = (500, 100)$. The rationale for incorporating speed values when a violation occurs is to ensure that violations occurring at higher speeds are assigned higher violation scores.



\subsection{Training and Testing}
Our experiments are divided into two phases: training and testing, each aimed at evaluating the effectiveness of our framework. 

\subsubsection{Training}
The training experiments were conducted in CARLA's ~\cite{dosovitskiyCARLAOpenUrban2017b} \textit{Town10}, an urban map setting with multiple four-way intersections. To ensure the DRL agent experiences a broad range of initial scene configurations, we start each episode by randomly selecting from 1202 pre-cataloged scenes, each featuring vehicles with randomized colors and makes. This setup is further enhanced by the stochastic behaviors of non-ego vehicles, governed by CARLA's~\cite{dosovitskiyCARLAOpenUrban2017b}  traffic manager, introducing both realism and complexity. The variable vehicle arrangement, appearances, and the other vehicle agents' stochastic nature significantly enriched the training landscape, equipping the DRL agent to adeptly uncover the system's vulnerabilities. The DRL agent is trained with 40960 steps, with an episode length of 512. The policy of the DRL agent updates every four trajectories.

\subsubsection{Testing}
The testing experiments unfolded within CARLA's~\cite{dosovitskiyCARLAOpenUrban2017b} \textit{Town05}, a map setting featuring pine-covered hills and a network of roads and a highway. In this phase, we aimed for diverse yet repeatable conditions by selecting 50 out of 1199 cataloged scenes from the initial scene generator. This is realized by using a fixed random seed, ensuring these scenarios are consistent across different experiment runs. The deterministic setting extends to other vehicles' appearances and their behavior, i.e., the color/make of the vehicles and the path they follow throughout the testing episode are deterministic, chosen to maintain uniformity to allow each scenario's outcomes to be directly comparable. This method ensures that despite the inherent variability of the test runs, the foundational conditions remain constant, facilitating an accurate assessment of performance across varying scenarios.

Records of violation scores and other information were kept, with averages calculated for analysis. These findings were benchmarked against two specific scenarios: our system navigating in clear weather and under randomly perturbed weather conditions by a non-strategic agent, providing a comprehensive evaluation of the GENESIS-RL framework. The length of each testing episode is identical to that of the training, which is 512 time steps in the simulation.


\subsection{Evaluation Metrics}
In our evaluation process, we employ two metrics: the violation score and the minimum following distance deficit $\delta_{mfd}$. The former metric has been detailed earlier in this section and measures the system's adherence to a set of predefined safety rules. For the latter, we introduce a metric by leveraging the concept of minimum following distance $d_{min}$, as defined in Responsibility-Sensitive Safety (RSS) \cite{hasuo2022responsibility}. We adopt a simplified version of the RSS criterion, which assumes that both the ego and lead vehicles have the same maximum deceleration rate and that the reaction time of the autonomous system is negligible. The simplified formula for calculating $d_{min}$ is given by the following equation:
\begin{align}
    d_{min} = \max\{0, (\frac{v_{e}^2-v_{l}^2}{2a})\}
\end{align}
where $v_e$, $v_l$ the speed of the ego and lead vehicle, respectively; $a$ the maximum deceleration of both vehicles, which is 5 $m/s^2$. 

The minimum following distance deficit $\delta_{mfd}$ is then calculated by the discrepancy between the actual distance $d$ maintained by the ego vehicle and the calculated $d_{min}$, i.e., $\delta_{mfd} = \max(0, d_{min} - d)$, where $d$ is the distance between the ego and lead vehicle. Specifically, it quantifies how much closer the ego vehicle comes to the lead vehicle than is deemed safe according to the simplified RSS criterion. In the following section, we will show the sum of this metric over all 50 runs. Showing the extent of the ego vehicle getting too close to the lead car. 

\subsection{Computational Time}
The experiments were conducted on a computational setup with a 32-core CPU, 64GB of RAM, and an NVIDIA GTX TITAN X graphics card. A complete training run required approximately 3-4 hours on this hardware, while a testing run took about 1-2 minutes per episode. 

\section{Results and Discussion}
In this section, we present the outcomes of our experiments, including an analysis of the performance and efficacy of our framework. The results underscore the capability of our approach to generate meaningful edge cases, highlighting key findings and insights gained through testing scenarios. We validate our approach by answering the following questions:
\begin{itemize}
    \item What impact do the generated edge cases have on the performance and decision-making processes of the system? 
    \item How effectively does our framework generate edge cases that are both challenging and realistic for the system under test?
\end{itemize}


\subsection{Impact of GENESIS-RL Generated Edge Cases on System Performance}

\begin{figure}[ht]
\includegraphics[width=\linewidth]{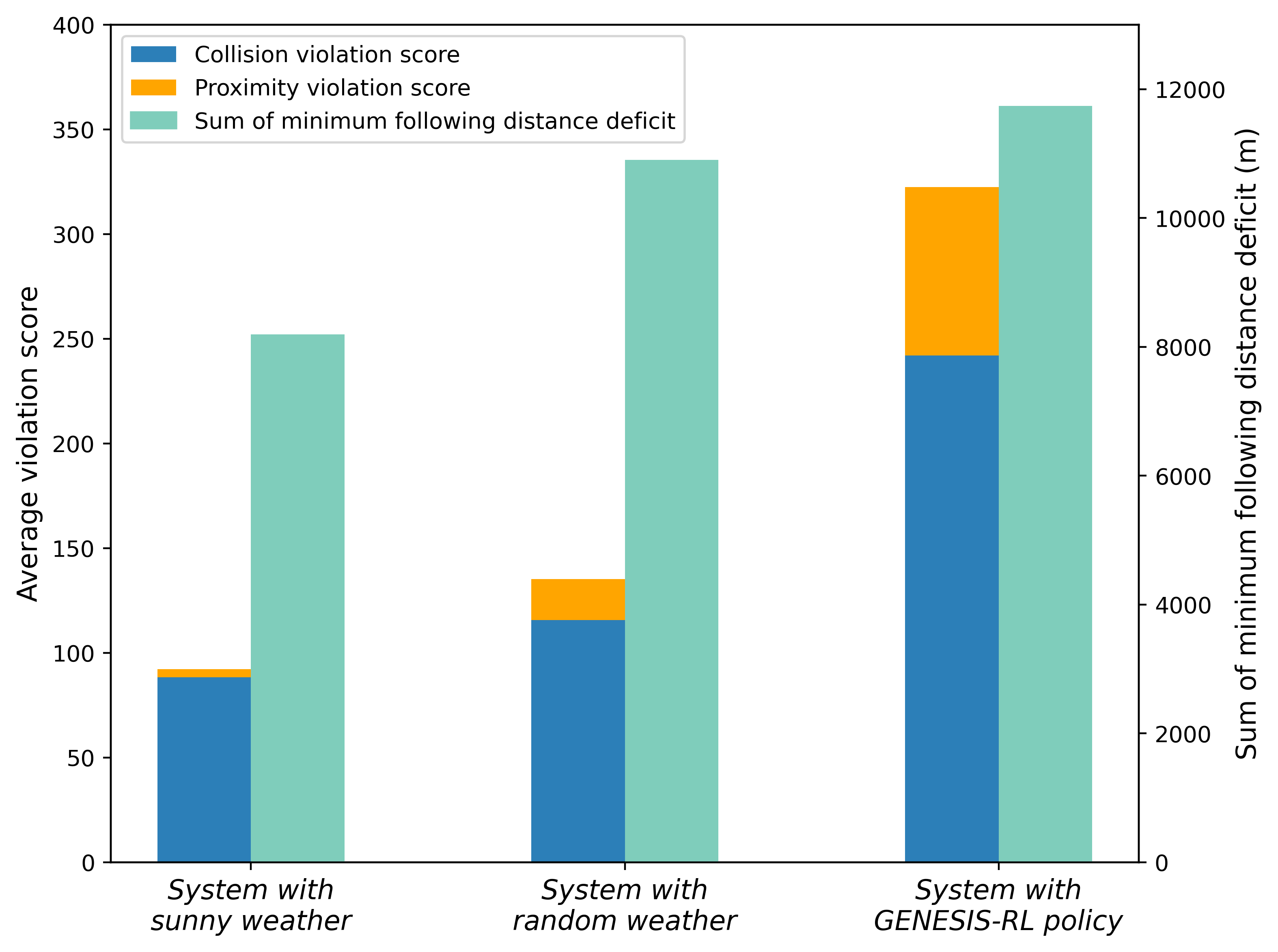}
\caption{Violation scores and the sum of minimum following distance deficit (based on RSS) across three testing scenarios - the system operates under sunny weather, random weather and under the GENESIS-RL policy. The results presented are averages across 50 runs with randomly selected initial scenes.}
\label{fig:violation-mindist2stop}
\end{figure}

Our analysis of GENESIS-RL's impact on the system reveals a notable increase in both violation scores (See Fig.~\ref{fig:violation-mindist2stop}) as well as the sum of minimum following distance deficit, illustrating its ability to effectively challenge and exploit system vulnerabilities. Specifically, the scenario output from GENESIS-RL's policy leads to significant variations in system performance, as demonstrated by the following observations:

\subsubsection{Proximity violation score increase} Under weather scenarios controlled by GENESIS-RL, the system's braking response was markedly delayed compared to its reaction under other testing scenarios and, in some instances, the braking behavior is altogether absent. This delay/absence is quantifiably demonstrated through the analysis of the ego vehicle's telemetry data (See Fig.~\ref{fig:timeseries}(c), the system operating under the GENESIS-RL policy maintains brakes much later, i.e., the blue curve is shifted more to the right compared to the two other scenarios), showcasing a contrast in the system's ability to maintain safe following distances under varied environmental influences.

\subsubsection{Collision violation score increase}
\begin{figure}[htb]
\includegraphics[width=\linewidth]{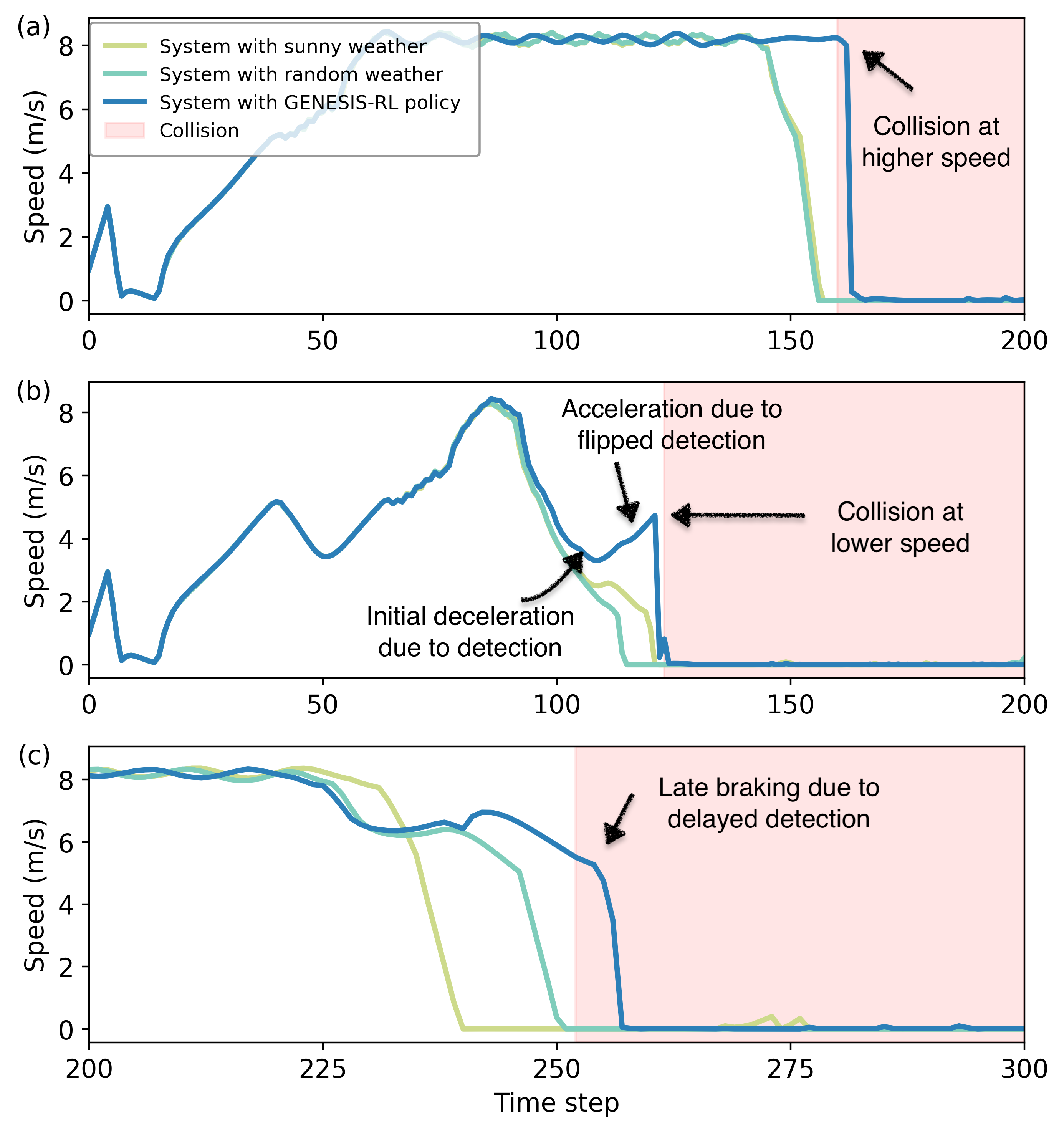}
\caption{Examples of the system failure modes based on vehicle telemetry. (a) Example of a non-detection collision - where the system crashed into the vehicle in front at full/high speed due to the non-detection of the other vehicle. (b) Intermittent detection collision - where the intermittent detection prevents the ego vehicle from stopping in time, leading to a lower-speed collision. 
(c) Delayed detection collision - where the system detects the lead car too late to stop in time. In the tests depicted in this figure, the ego vehicle successfully avoided collisions across both sunny and random weather scenarios and only failed under the conditions generated by GENESIS-RL. 
 }
\label{fig:timeseries}
\end{figure}

Further looking into the types of collision violations induced by GENESIS-RL, we categorize them into three main failure modes:
\begin{itemize}
    \item Non-detection collisions: Instances where the system fails to detect the leading vehicle at all, resulting in high-speed impacts (See Fig.~\ref{fig:timeseries}(a)). This failure mode underscores critical perception system vulnerabilities under complex environmental conditions orchestrated by GENESIS-RL.
    \item Intermittent detection collisions: Occurrences where the system initially detects the lead vehicle but subsequently loses track of it, leading to collisions at reduced speeds (See Fig.~\ref{fig:timeseries}(b). These incidents underscore the deficiencies in the system's ongoing tracking, and/or highlight the shortcomings in its ability to respond promptly within the scenarios induced by GENESIS-RL.
    \item Delayed detection collision: Occurrences where the system detects the lead vehicle too late to stop in time, leading to collisions at reduced speeds (See Fig.~\ref{fig:timeseries}(c). These incidents highlight the deficiencies in the system's detection mechanisms.
    
\end{itemize}

These findings not only exemplify GENESIS-RL's capability in uncovering and leveraging system weaknesses but also emphasize the imperative need for bolstered system robustness against a wide spectrum of real-world conditions. 

\subsection{Effectiveness of GENESIS-RL in Generating Edge Cases}

Our results show that with the GENESIS-RL framework, we were able to generate edge cases that pose significant challenges not only to automated systems but also to human perception and response capabilities. A prime example of such conditions includes scenarios combining foggy weather with heavy rainfall or nocturnal settings accentuated by heavy rain, where the reflections on wet surfaces severely disrupt the detection capabilities of autonomous systems. These conditions, inherently challenging due to their impact on visibility and sensor efficiency, highlight the scenario generation capabilities of GENESIS-RL, underscoring its potential for creating diverse testing environments that closely mimic real-world driving complexities. 

During inference, we observed that the DRL agent relies heavily on manipulating certain parametric knobs (such as rain and fog density) to introduce challenges to the system. To investigate GENESIS-RL's capability to generate a variety of edge cases, we conducted experiments with fog and rain density levels set to zero, individually and in combination. Despite the absence of these parametric knobs, our framework is still capable of creating edge cases that undermine the system's performance, demonstrating its robustness in edge case generation beyond the reliance on certain powerful and effective parametric knobs (See banner figure on the first page for reference).

\section{Conclusion and Future Work}
In conclusion, our study demonstrates the GENESIS-RL framework's capability to generate complex and challenging edge cases for autonomous systems, which are critical for thoroughly testing and enhancing the reliability of systems as such. Moreover, GENESIS-RL proved capable of producing edge cases that will lead to system failure even in the absence of some dominating factors, underscoring its robustness and potential broader application in safety-critical testing environments.

The implications of the results from our work is manifold. First, they highlight the need for including a wide range of challenging scenarios in the testing protocols for autonomous systems, ensuring they are well-equipped to handle the intricacies of complex and dynamic environments. Additionally, the ability of GENESIS-RL to generate test conditions without relying exclusively on dominant factors showcases its value in crafting safer, more dependable autonomous system solutions. 


In future work, we aim to expand our exploration to include a wider spectrum of parameters, such as a more comprehensive range of weather parameters, behaviors of other actors (e.g., vehicles and pedestrians) to further enhance the scenario generation capabilities of GENESIS-RL, thereby facilitating the creation of more diverse and challenging testing environments for autonomous systems.


\bibliographystyle{unsrt}
\bibliography{ref}
\vspace{12pt}

\end{document}